\documentclass[sigconf]{acmart}
\usepackage{balance}

\AtBeginDocument{%
  \providecommand\BibTeX{{%
    \normalfont B\kern-0.5em{\scshape i\kern-0.25em b}\kern-0.8em\TeX}}}

\setcopyright{acmcopyright}
\copyrightyear{2021}
\acmYear{2021}
\acmDOI{10.1145/nnnnnnn.nnnnnnn}

\acmConference{DLP-KDD 2021}{August 15, 2021}{Singapore} 
   
\acmBooktitle{3rd Workshop on Deep Learning Practice for High-Dimensional Sparse Data with KDD 2021, August 15, 2021, Singapore}
\acmPrice{15.00}
\acmISBN{978-x-xxxx-xxxx-x/YY/MM}

\begin{document}

\title{Neighbor Based Enhancement for the Long-Tail Ranking Problem in Video Rank Models}




\author{Ziyu He}
\authornote{This work was done during Ziyu’s internship at Tencent. Both authors contributed equally to this research.}
\email{davidzyhe@tencent.com}
\author{Xuanji Xiao}
\authornotemark[1]
\authornote{Xuanji Xiao is the Corresponding author.}
\email{growj@126.com}
\affiliation{%
  \institution{Tencent Inc}
  \streetaddress{Xigema building,Zhichun Road, Haidian qu}
  \city{Beijing}
  \country{China}
  \postcode{100080}
}

\author{Yongyu Zhou}
\affiliation{%
	\institution{Beijing Institute of Technology}
	\city{Beijing}
	\country{China}}
\email{zhouyongyu@bit.edu.cn}


\begin{abstract}
Rank models play a key role in industrial recommender systems, advertising, and search engines. 
Existing works utilize semantic tags and user-item interaction behaviors, e.g., clicks, views, etc., to predict the user interest and the item hidden representation for estimating the user-item preference score.
However, these behavior-tag-based models encounter great challenges and reduced effectiveness when user-item interaction activities are insufficient, which we called "the long-tail ranking problem". Existing rank models ignore this problem, but it's common and important because any user or item can be long-tailed once they are not consistently active for a short period. In this paper, we propose a novel neighbor enhancement structure to help train the representation of the target user or item. It takes advantage of similar neighbors (static or dynamic similarity) with multi-level attention operations balancing the weights of different neighbors. Experiments on the well-known public dataset MovieLens 1M demonstrate the efficiency of the method over the baseline behavior-tag-based model with an absolute CTR AUC gain of 0.0259 on the long-tail user dataset.\footnote{The source code: https://github.com/neighbourbasedrec/Neighbour\_Enhanced\_DNN.}
\end{abstract}





\maketitle
\section{Introduction}
Rank models, such as click-through rate prediction (CTR), post-click conversion rate prediction (CVR), are fundamental modules in industrial recommender systems (RS), advertising, and search engines. It can precisely predict user preference scores for items. However, there exists a well-known cold-start problem \cite{schein2002methods} that when users/items are new to the platform with few activities such as clicks and purchases it is difficult to provide accurate predictions. On the other hand, many low-frequency or long-tail users/items have similar experience to cold start as a result of lack of activities in a relatively recent period. Given that users/items may not be active all the time, any user/item has a random opportunity to become a long-tail user/item sometimes. We uniformly summarize these two problems in rank models as \textbf{"the long-tail ranking problem"}. Undoubtedly, it is both challenging and promising to tackle it.
\begin{figure}[htp]
 \begin{center}
   \vspace{-10pt}
  \includegraphics[width=\linewidth]{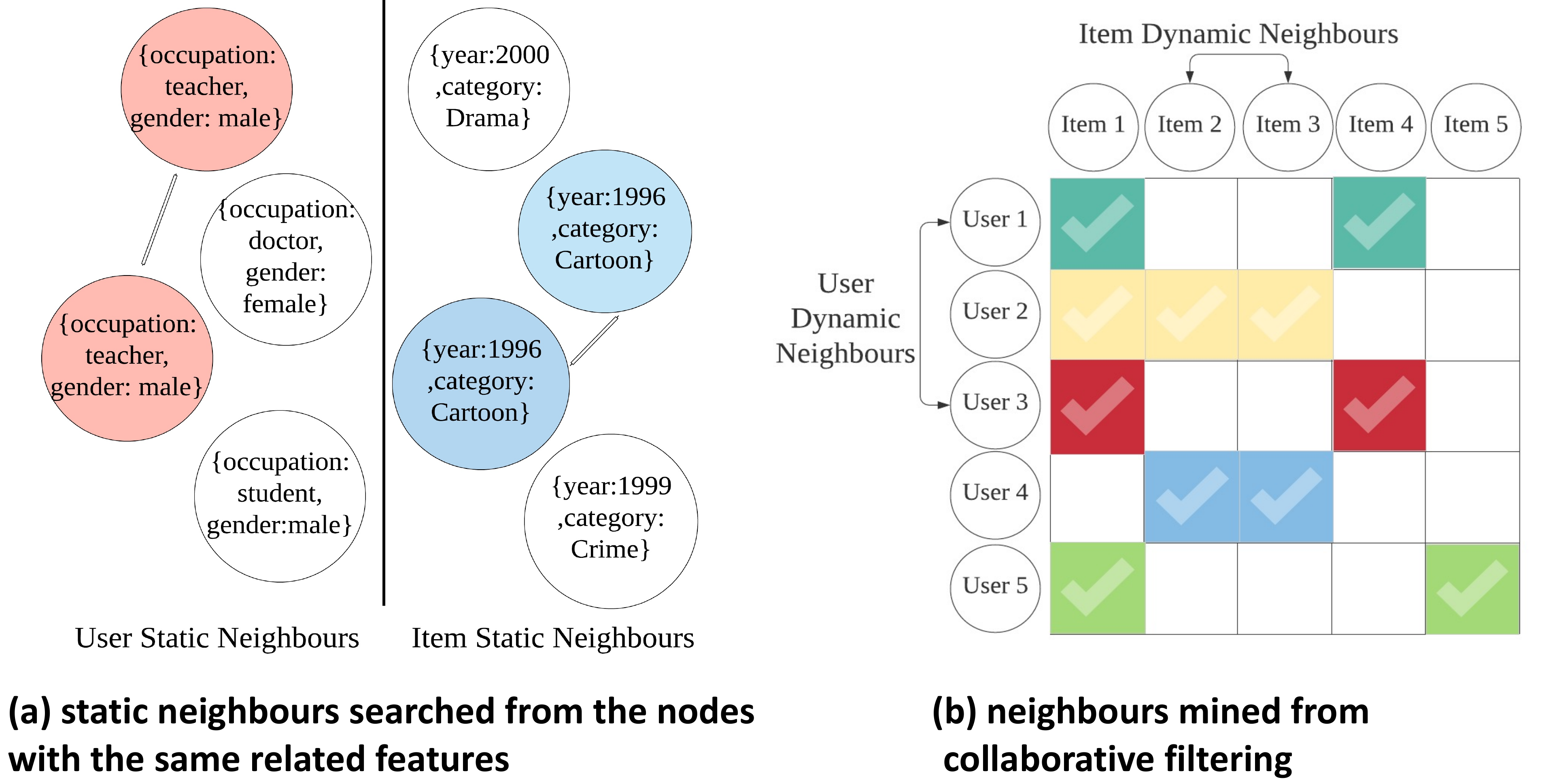}
  \end{center}
  \vspace{-15pt}
  \caption{static and dynamic neighbors}
  \vspace{-10pt}
  \label{figure1}
\end{figure}

The cold start situation has been explored in the match models \cite{wang2018billion,grbovic2018real} in RS but almost unnoticed by the research field of rank
models. The match model, as the pre-step of the rank model, is responsible for initially selecting a batch of items, from which the rank model further selects the best items for the user \cite{cheng2016wide}. In the field of match algorithms, EGES~\cite{wang2018billion} proposes to merge side information with item embeddings generated through a graph embedding model and considers the importance of different side information. Grbovic et al.~\cite{grbovic2018real} propose to create embeddings for new listings by utilizing existing embeddings of other listings similar to them. 
However, current rank models~\cite{covington2016deep, cheng2016wide, zhou2018deep, zhou2019deep} mainly concentrate on semantic properties (user demographic properties and item semantic tags) and user-item interaction behaviors (clicks, views, etc.) to predict the user interest and item hidden representation. Among them, YoutubeDNN \cite{covington2016deep} combines the embedding representation of user/item with a user historical behavior sequence, which could well capture some extent of user/item relations. DIN and DIEN~\cite{zhou2018deep, zhou2019deep} models modify the GRU temporal model to extract the evolving user interest behind historical clicks. 
Obviously, they have been indulged by the traditional reliance on sufficient historical click activities for a given user or item to train its representation, while on long-tail conditions there is none to depend on.

 Therefore, to solve the long-tail ranking problem mentioned above, we first propose a neighbor enhancement structure that can help to train and enhance the representations of target users/items by aggregating the embeddings of their neighbors. Furthermore, we combine the neighbor enhancement structure with existing rank models to strengthen their ability dealing with long tail users/items. Generally, it is assumed that similar users favor similar items, and therefore the representations of long-tail users/items could benefit from their similar neighbors.

We refer to users/items as nodes in the following paper. In our model, two types of neighbors can be obtained to enhance the representation of the target node (Fig.\ref{figure1}). The first type is the \textbf{static} neighbors sharing the same semantic tags of a given target node (Fig.\ref{figure1}(a)). We assume that the nodes with the same tags should possess similar long-term interests. The second type is the \textbf{dynamic} neighbors mined from user-item interaction matrix by collaborative filtering techniques~\cite{paterek2007improving} (Fig.\ref{figure1}(b)). The assumption regarding the dynamic neighbors is that the neighbors of the target node exhibit similar click behaviors. Lastly, the neighbor enhancement structure, consisting of two layers of network attention structures, is designed to aggregate and balance the embeddings of these two types of neighbors to form the final representation of the target node. 

To summarize our main contributions:
\begin{itemize}
\item
We design a novel neighbor-based method to enhance the user/item representation for tackling the long-neglected "long-tail ranking problem" in traditional behavior-tag-based rank models. The method first obtains the neighbors of the target user/item from two perspectives (static properties/dynamic interest), and then uses multi-level attention to aggregate neighbors as part of the user/item representation. 
\item
We conducted extensive experiments on the public dataset MovieLens 1M, and the results demonstrate the efficiency of our method in solving the long-tail ranking problem with a significant AUC gain of 0.0259 in the long-tail MovieLens 1M' sparse dataset.
\item
As far as we know, this method is the first to explicitly consider the "long-tail ranking problem" in rank models. Due to the model independence of this method, it can be easily integrated with any other existing state-of-the-art rank models to improve their performance.
\end{itemize}

\begin{figure*}[h]
  \begin{center}
  \includegraphics[width=1\textwidth,height=0.31\textheight]{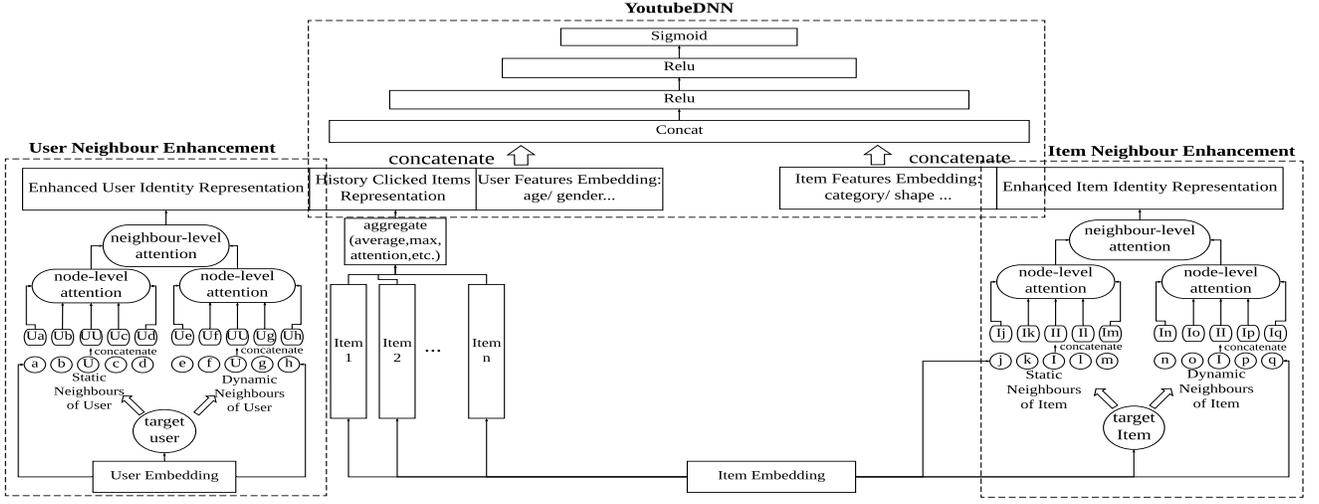}
  \end{center}
  \vspace{-15pt}
  \caption{Neighbor-Enhanced-YoutubeDNN model architecture. The user neighbor enhancement and item neighbor enhancement are on the left and right sides of the classic YoutubeDNN structure respectively. The enhanced user/item representation is concatenated with other embeddings and then fed into the DNN model to output the predicted value.
  }
    \vspace{-15pt}

  \label{main}
\end{figure*}

\section{The proposed approach}
In this section, we propose in detail our method (Fig.\ref{main}), along with its core components including searching neighbors and neighbors aggregation. We choose YoutubeDNN as the baseline model because YoutubeDNN is proven effective on the real-world video recommendation platform Youtube \cite{covington2016deep}, which is similar to our situation. 


\subsection{Searching Neighbors of the Target Node}
As shown in Figure~\ref{figure1}, given the target node, we can obtain the \textbf{static} neighbors by searching the neighbors with same characteristics, from which we drive the semantic connection (long term similarity). We can obtain the \textbf{dynamic} neighbors by mining from their historical click interactions to extract similarity in behavior (short time similarity). Specifically, we choose all the other nodes with the same static attributes of the target node as the static neighbors, and we calculate the cosine similarity between the users and items vectors in the user-item interaction matrix and then choose the most similar nodes as the dynamic neighbors. For example, regarding user 1, aged 20 and taking up the occupation as a teacher, we can find neighbors with the two identical attributes. In the dynamic situation, user 1 clicked item 1 and item 4, then we can find his neighbors who also clicked item 1 and item 4 (Fig. \ref{figure1}). The process of finding neighbors is similar for that of items. 
The static neighbors searching can be formulated as meta-path \cite{sun2012mining}:
\begin{equation}
	User_{1} \stackrel{age}{\longrightarrow} 20 \stackrel{age}{\longrightarrow}User_{2}
\end{equation}
\begin{equation}
	User_{1} \stackrel{occupation}{\longrightarrow} teacher \stackrel{occupation}{\longrightarrow}User_{2}
\end{equation}
We will choose the node with most metapaths connecting the target node as the static neighbor. The dynamic neighbors searching can be formulated as the vector distance \cite{paterek2007improving}:
\begin{equation}
\operatorname { sim } ( target , candidate ) 
= \frac { \vec { target } \cdot \vec { candidate } } { \| \vec { target } \|  * \| \vec { candidate } \|  }
\end{equation}
The nearest node will be chosen as the dynamic neighbors. The static and dynamic neighbor searching process can be tuned through more complex and refined calculation methods \cite{sun2012mining,paterek2007improving}.

\subsection{Node-level Aggregation}
\begin{figure}[h]
  \centering
  \includegraphics[height=6cm]{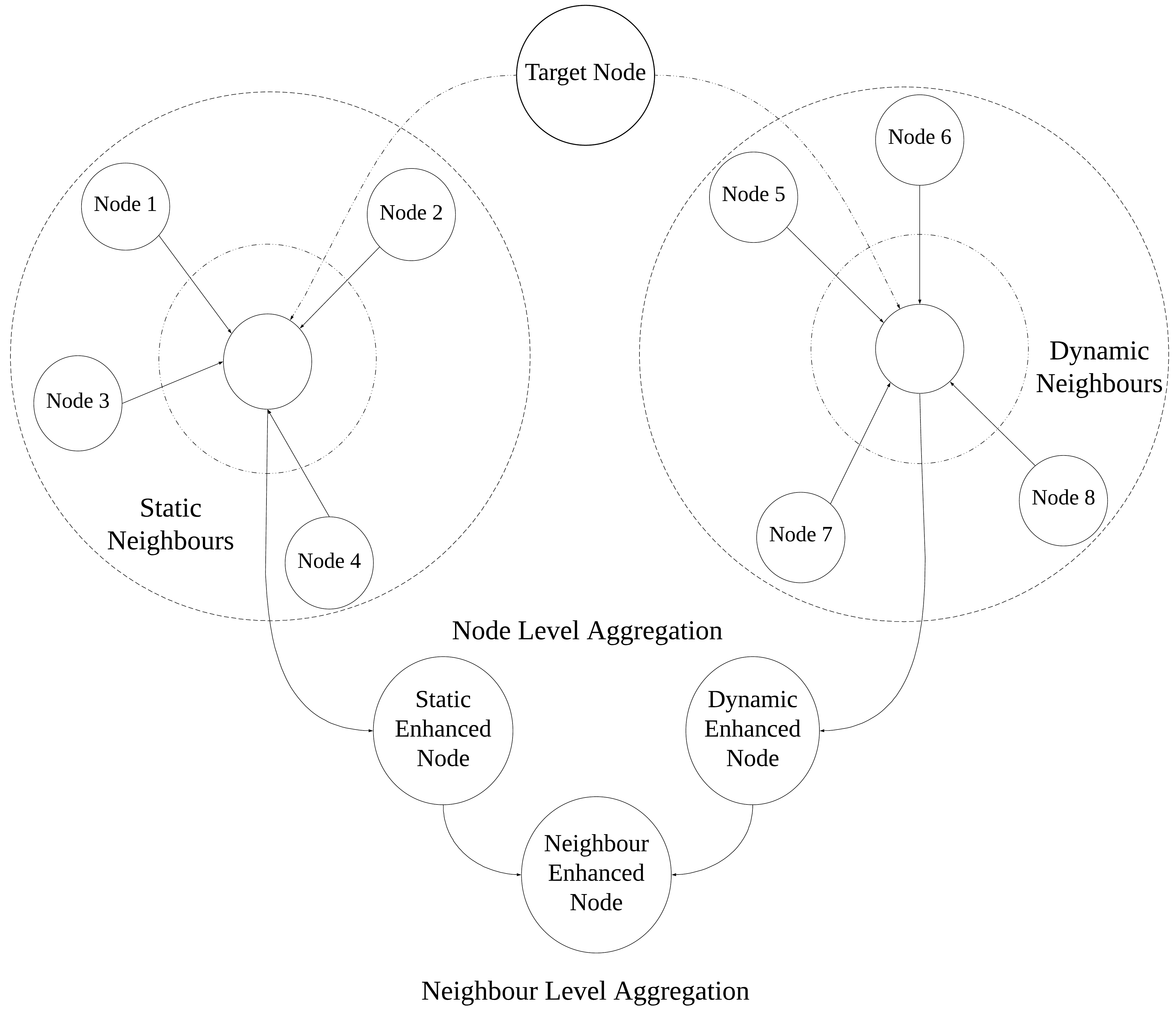}
  \vspace{-10pt}
  \caption{The two-level aggregation of neighbor enhancement structure}
  \label{figure3}
    \vspace{-22pt}
\end{figure}

The node-level attention layer (Fig.\ref{figure3} and Fig.\ref{main}) is designed to aggregate the embeddings of the neighbors of a given node within each type of neighbors(\textbf{static} or \textbf{dynamic}). We aggregate the neighbor nodes embeddings and the target node embedding together to form a neighbor-specific representation of the target node by leveraging the graph attention network(GAT)~\cite{velivckovic2017graph} structure, which takes into account the importance of each neighbor to the target node. Both users and items could utilize this structure to enhance their identity representations. 

Regarding each type of neighbor relation, we firstly derive the weight of each neighbor node $h_{j}$ to the target node $h_{i}$ within the neighbor relation $\theta$ (\textbf{static} or \textbf{dynamic}), denoted as $e_{i, j}^{\theta}$, by the following equation:
\begin{equation} \label{eqn}
 e_{i, j}^{\theta} = q_{\theta}^T \times [Wh_i || Wh_j],
\end{equation}
where $q_{\theta}^T$ is the learned attention variable that is shared among all neighbor-target pairs. $W$ is the trainable linear transfer matrix. We normalize all the weights of the node pairs $e_{i, j}^{\theta}$ of the neighbor nodes $j \in N_i^\theta$ to the target node $i$ to get the coefficient of the node pair weight $a_{ij}^{\theta}$ via softmax function:
\begin{equation} \label{eqn}
 a_{ij}^{\theta} = \frac{exp(e_{i, j}^{\theta})}{\sum_{k \in N_i^\theta}exp(e_{i, k}^{\theta})}.
\end{equation}

Finally, we take a weighted sum of all the neighbor embeddings to get the final representation of the target node within the neighbor relation $\theta$:
\begin{equation} \label{eqn}
z_i^{\theta} = \sigma(\sum_{j \in N_i^\theta} a_{ij}^{\theta} \times h_j),
\end{equation}
where $z_i^{\theta}$ represents the aggregated representation of target node $i$ within the neighbor relation $\theta$, and the $\sigma$ represents the ReLU activation function.

\subsection{Neighbor-Type-Level Aggregation}
As mentioned above, there are two types of neighbors involved in our model, static and dynamic (Fig.\ref{figure3}), which reflect two perspectives of enhancement. The static neighbors possess the same characteristics as the target node. The dynamic neighbors exhibit similar click behaviors to the target node, and therefore they have similar click interests. Some users may have constantly changing click interests, for whom the static neighbor representation could be more important. For the users with stable click interests, their dynamic neighbor representation may be more significant. Therefore, we take into account the different importance of each merged neighbor-specific representation and propose a neighbor-type-level aggregation layer to aggregate the two merged representation $z_i^{\theta}$ derived from the node-level attention to form the final representation $Z_i$ of the target node. The importance of each neighbor-specific representation $w_{\theta_i}$ is calculated as follows:
\begin{equation} \label{eqn}
    w_{\theta_i} = q^{T} \times tanh(W \times z_i^{\theta} + b),
\end{equation}
where $q \in R^{F'}$ is a learnt query attention variable, and the $W \in R^{F \times F'}$ and $b \in R^{F'}$ is trained to conduct a non-linear transformation of the input. After this, the $w_{\theta_i}$ is normalized by a softmax function to obtain the weight of each neighbor-specific representation (\textbf{static} or \textbf{dynamic}):
\begin{equation} \label{eqn}
    \alpha_{\theta_i} = \frac{exp(w_{\theta_i})}{\sum_{i = 1}^{2}exp(w_{\theta_i})}.
\end{equation}
Finally, we obtain the final representation of the target node by taking the weighted sum of two neighbor-specific representations:
\begin{equation} \label{eqn}
    Z = \sum_{i = 1}^{2}\alpha_{\theta_i} \times z_i^{\theta}.
\end{equation}
 The final representation $Z$ of the target node is a weighted combination of static and dynamic neighbor-specific representation. Therefore, the merged final representation of the target node could be more comprehensive than the representation of the target node itself. Logically, this neighbors aggregation effect is more significant for the representations of long-tail users/items. 

\subsection{Neighbor Enhanced YoutubeDNN }
As shown in Figure \ref{main}, The embeddings of the historical click items are aggregated by averaging to obtain a dynamic representation of user click interests. The embeddings of different feature groups including the representations of the enhanced user and item identity are concatenated together and fed into the multi-layer perceptrons (MLPs). Therefore, the predicted click-through rate(CTR) value could benefit from the neighbor enhancement structure, because more sufficient representations of users and items could be obtained. 

\section{Experiments}
\subsection{Experimental Setup}
\textbf{Datasets.}
The experiments are conducted on the public movie rating dataset MovieLens 1M\footnote{\url{https://grouplens.org/datasets/movielens/1m/}}. To demonstrate the effects of our model on mitigating the long-tail ranking problem, we have conducted some preprocessing to produce two variations of this dataset of different extent of long-tail. Table~\ref{tab:commands} summarizes the statistics of the datasets.

\textbf{\emph{MovieLens 1M.}}
The dataset consists of 1 million movie ratings given by 6040 users to 3883 movies. However, each user has at least 20 movie ratings, which means there are no long-tail users. 

\textbf{\emph{MovieLens 1M'.}}
We randomly select 1/10 of all the historical interactions of the \emph{MovieLens 1M} dataset to generate long-tail users and items, denoted as MovieLens 1M'. 1558 users have rated less than 5 movies, taking account of 26\% of all users.

\textbf{\emph{MovieLens 1M' Sparse.}}
We further process \emph{MovieLens 1M'} dataset by removing users with more than 30 historical interactions, which produces a dataset with a higher proportion of long-tail users and items, with 31\% users having less than 5 movie ratings.

To simulate the online CTR prediction problem, we consider all the rating instances given by users to movies as the positive click instances, whose label is 1. We randomly sample the other movies never viewed by the user as the negative click samples, labeled as 0. We randomly select 3 negative instances for each positive instance. 

\begin{table}
  \caption{MovieLens 1M, 1M', and 1M' Sparse datasets}
  \label{tab:commands}
  \vspace{-10pt}
  \begin{tabular}{cccl}
    \toprule
     & Users & Items & Interactions\\
    \midrule
    MovieLens 1M & 6040 & 3883 & 1000209\\
    MovieLens 1M' & 5973(99\%) & 3333(86\%) & 100021\\
    MovieLens 1M'Sparse & 4956(82\%) & 2905(75\%) & 46992\\
    \bottomrule
  \end{tabular}
  \vspace{-10pt}

\end{table}


\textbf{Model Comparison.}
We choose the state-of-art video rank model YoutubeDNN as our baseline to verify the usefulness of our method in solving the long-tail ranking problem. Note that the neighbor enhancement structure can be easily integrated with any other rank models, e.g., DIN or DIEN, because it doesn't interfere with the upper MLP structure (Fig.\ref{main}).

\textbf{Parameter Settings.}
(1) neighbors setting. The number of neighbors of the target node using in the final model (Fig.\ref{main}) is a hyperparameter, which can be tuned manually. In our case, we choose at most 50 static neighbors and 10 dynamic neighbors for each target node.
(2) network setting. The embedding size of the user/item ID: 32, the network structure: 128-64-32, the optimizer: Adam, the activation function: ReLU in the first two layers and sigmoid in the last layer, dropout rate: 0.3, learning rate:0.005. These settings are shared between our Neighbor-Enhanced-YoutubeDNN model and the baseline YoutubeDNN model. For the neighbor enhancement structure, the dimensions of the q attention variable of the two attention layers are set to be 64 and 32, respectively.
\subsection{Results}

\begin{table}
  \caption{Experiment result on MovieLens 1M}
  \label{tab.1m}
  \vspace{-10pt}
  \begin{tabular}{cl}
    \toprule
     Model&AUC \\
    \midrule
    YoutubeDNN & 0.9779 \\
    Neighbor-Enhanced-YoutubeDNN& 0.9790(+0.0011)\\
    \bottomrule
  \end{tabular}
    \vspace{-10pt}
\end{table}

\begin{table}
  \caption{Experiment result on MovieLens 1M'}
  \label{tab.1mm}
  \vspace{-10pt}
  \begin{tabular}{cl}
    \toprule
     Model&AUC \\
    \midrule
    YoutubeDNN & 0.9579 \\
    Neighbor-Enhanced-YoutubeDNN & 0.9631(+0.0052)\\
    \bottomrule
  \end{tabular}
  \vspace{-10pt}
\end{table}

\begin{table}
  \caption{Experiment result on MovieLens 1M' Sparse}
  \label{tab.1mmsparse}
  \vspace{-10pt}
  \begin{tabular}{cl}
    \toprule
     Model&AUC \\
    \midrule
    YoutubeDNN & 0.8872 \\
    Neighbor-Enhanced-YoutubeDNN & 0.9131(+0.0259)\\
    User-Enhanced-YoutubeDNN & 0.9081(+0.0218) \\
    Item-Enhanced-YoutubeDNN & 0.8913(+0.005) \\
    \bottomrule
  \end{tabular}
\vspace{-15pt}
\end{table}


Overall, the experimental results of the enhanced model surpass the baseline model in all three datasets. We use the $AUC$ value (Area Under ROC) as the evaluation metric for the CTR experiments. An AUC gain of 0.001 is remarkable, which could lead to a significant online performance improvement in industrial recommendation systems. In MovieLens 1M dataset, it increases 0.0011 in absolute AUC value, which indicates that even if the situation is not long-tailed there may also get an improvement (Table~\ref{tab.1m}). For those two long-tail datasets, the advantages are obvious, with absolute AUC gains rising to 0.0052 and 0.0259 (Table~\ref{tab.1mm} and Table~\ref{tab.1mmsparse}). Besides, we conducted ablation tests in the MovieLens 1M' Sparse dataset to demonstrate the effect of neighbor enhancement structure from user and item perspectives separately. The User-Enhanced-youtubeDNN and Item-Enhanced-youtubeDNN in Table~\ref{tab.1mmsparse} represent the models that include the neighbor enhancement structures only in user perspective and item perspective respectively. The absolute AUC gains of them are about 0.0218 and 0.005. It confirms the assumption that the neighbor enhancement structure can enhance the representation of both users and items, mitigating the long-tail 
ranking problem.

\section{CONCLUSIONS AND FUTURE WORK}
In this paper, we tackle the long-neglected "long-tail ranking problem" with a novel neighbor enhancement method. We obtain the neighbors of the target user/item from two perspectives, static and dynamic, and then enhances the representation of a given user/item by aggregating its neighbors with multi-level attention, which weights different neighbor nodes and balance the effects of static and dynamic neighbors. Public dataset experiments demonstrate the effectiveness of our suggested method.
In the future, we will try more advanced neighbor searching methods to further mitigate "the long-tail ranking problem" and explore the algorithm effectiveness in other rank applications, such as CVR, like rate. 

\bibliographystyle{ACM-Reference-Format}
\balance
\bibliography{sample-base}

\end{document}